\documentclass[aps,prl,reprint,longbibliography]{revtex4-2}
\usepackage{graphicx}
\usepackage{amsmath}
\usepackage{amssymb}
\usepackage{color}
\usepackage[colorlinks,urlcolor=blue,citecolor=blue,linkcolor=blue]{hyperref}
\usepackage{gensymb}
\usepackage{array,multirow}

\begin{document}
\title{Magnetoelectric torque in polar magnetic bilayers}
\author{Zhong Shen}
\author{Jun Chen}
\author{Xiaoyan Yao}
\email[]{yaoxiaoyan@seu.edu.cn}
\author{Shuai Dong}
\email[]{sdong@seu.edu.cn}
\affiliation{Key Laboratory of Quantum Materials and Devices of Ministry of Education, School of Physics, Southeast University, Nanjing 211189, China}
\date{\today}

\begin{abstract}
Energy-efficient fast switching of spin orientations or textures is a core issue of spintronics, which is highly demanded but remains challenging. Different from the mainstream routes based on spin-transfer torque or spin-orbit torque, here we propose another mechanism coined as magnetoelectric torque to switch the magnetization in polar magnetic bilayers via pure electric field. In some magnetic van der Waals bilayers, when the electrostatic energy of polarization can compensate the interlayer magnetic coupling, a magnetoelectric torque is generated to fastly flip spins within a few picoseconds, which is demonstrated by combining the first-principles calculations, analytic model, as well as atomistic simulations. Such a magnetoelectric torque doesn't rely on the spin-orbit coupling and is generally active in polar magnetic homostructures and heterostructures. Our work provides an alternative route to switch magnetization in nanoscale, which may benefit the energy-saving and fast-response spintronic devices.
\end{abstract}
\maketitle

One core operation of spintronics is to switch spin orientations or textures between two distinct states fastly and efficiently. Currently, two mainstream techniques are being investigated: the relatively mature spin-transfer torque (STT)~\cite{Ralph2008JoMaMM} and the emerging spin-orbit torque (SOT)~\cite{Manchon2019RoMP}. However, both STT and SOT need the injection of charge current to generate spin current carrying corresponding torques. Thus, the Joule heat remains inevitable in the whole procedure and the energy consumption remains unsatisfactory even though it can be reduced somehow in principle.

From the aspect of multiferroicity, an alternative route is to switch spins via pure electric field ($E$), i.e., the converse magnetoelectricity. In principle, this approach is naturally more energy saving since the steady current is avoided. For example, based on BiFeO$_3$ films (and its derivants) as well as several ferroelectric (FE) ferromagnetic (FM) heterostructures, some pioneer works have demonstrated the possibility to switch the magnetization by electric field, via the FE field effects or Dzyaloshinskii-Moriya interaction (DMI)\cite{Dong2013PRB,Chen2021PRL,Yu2023PRL,Yu2024PRL}. In these cases where the ferroelectricity and magnetism own different origins, the FE ions' displacements play as the key messager, which limits their switching speeds due to the relatively slow motion of ions~\cite{Yu2024PRL,Yu2023PRL,Lin2017PRM}. In addition, there were also some efforts to switch the magnetization by pure electron redistribution \cite{Weng2016PRL,Lin2017PRM}, which can be faster in principle but remains unavailable in experiment so far.

Moreover, in those type-II multiferroics whose ferroelectricity originates from specific magnetic orders, their magnetoelectric coupling is intrinsically robust, but their polarizations ($P$'s) are typically faint (mostly $<0.1$ $\mu$C/cm$^2$) \cite{Kimura2003N,Dong2015AP,Fert2024RoMP}. Thus, the electrostatic energy is very small. For example, with an electric field $E=100$ kV/cm and $P=0.1$ $\mu$C/cm$^2$, the $E\cdot P$ is merely $4$ $\mu$eV/u.c. if the u.c. volume is $64$ \AA$^3$. For comparison, the magnetic coupling $J$ generally exceeds $1.0$ meV for most magnets. Thus, it may be easy to manipulate the ferroelectricity by magnetic field in these type-II multiferroics~\cite{Kimura2003N}, but rather difficult to switch their magnetism by electric field.

In recent years, two-dimensional (2D) van der Waals (vdW) magnets have evolved into a flourishing family, full of opportunities for the electrical field control of magnetism. An inborn advantage is that the electric field can be naturally large for a moderate voltage applied on these atomic-level layers. In addition, the magnetic coupling between vdW layers is naturally weak, making their magnetism more sensitive to external stimulations. For example, the electric-field-induced magnetization switching has been reported in CrI$_3$-based vdW systems \cite{Jiang2018NN,Jiang2018NM,Huang2018NN}. The underlying mechanism is the conventional electrostatic doping via gating field, which can tune the electronic structure and thus the interlayer magnetic coupling. However, the doping may damage the inherent insulation~\cite{Zhao2025S} and lead to the leakage current. Thus, it's vital and urgent to find other strategies for the electrical field control of magnetism in these vdW layers, and clarify the key physics such as spin dynamics.

In this Letter, an alternative mechanism named magnetoelectric torque (MET) is proposed to switch magnetization by electric field in polar vdW magnetic bilayers. Benefitting from its pure electronic origin, the magnetization can be fastly switched within a few picoseconds. Such MET does not rely on spin-orbit coupling (SOC), and thus generally works in polar magnetic bilayers. Although the terminology MET was once proposed in previous studies \cite{Zheng2017CPB,Xing2013SaAAP,Xing2011JoAP,Nan2018AFM,Sousa2021PRR}, our work is quite different from them (as discussed in End Matter). 

Here the Janus CrISe bilayer is taken as the model system. First, it's a derivative of the experimentally synthesized CrTe$_2$~\cite{Sun2020NR,Zhang2021NC}. Similar Janus structures have been synthesized in experiments~\cite{Lu2017NN,Trivedi2020AM}. Besides, its monolayer owns excellent dynamic stability and a relatively high Curie temperature~\cite{Shen2022APL,Zhu2023nCM}. Second, its plain FM ground state with strong magnetoelectricity provides an ideal platform to clearly demonstrate the physics of MET avoiding the obscurity caused by complex magnetic structures. Details of computational methods can be found in the Supplemental Materials (SM)~\cite{SMp3}.

\begin{figure}
\centering
\includegraphics[width=0.48\textwidth]{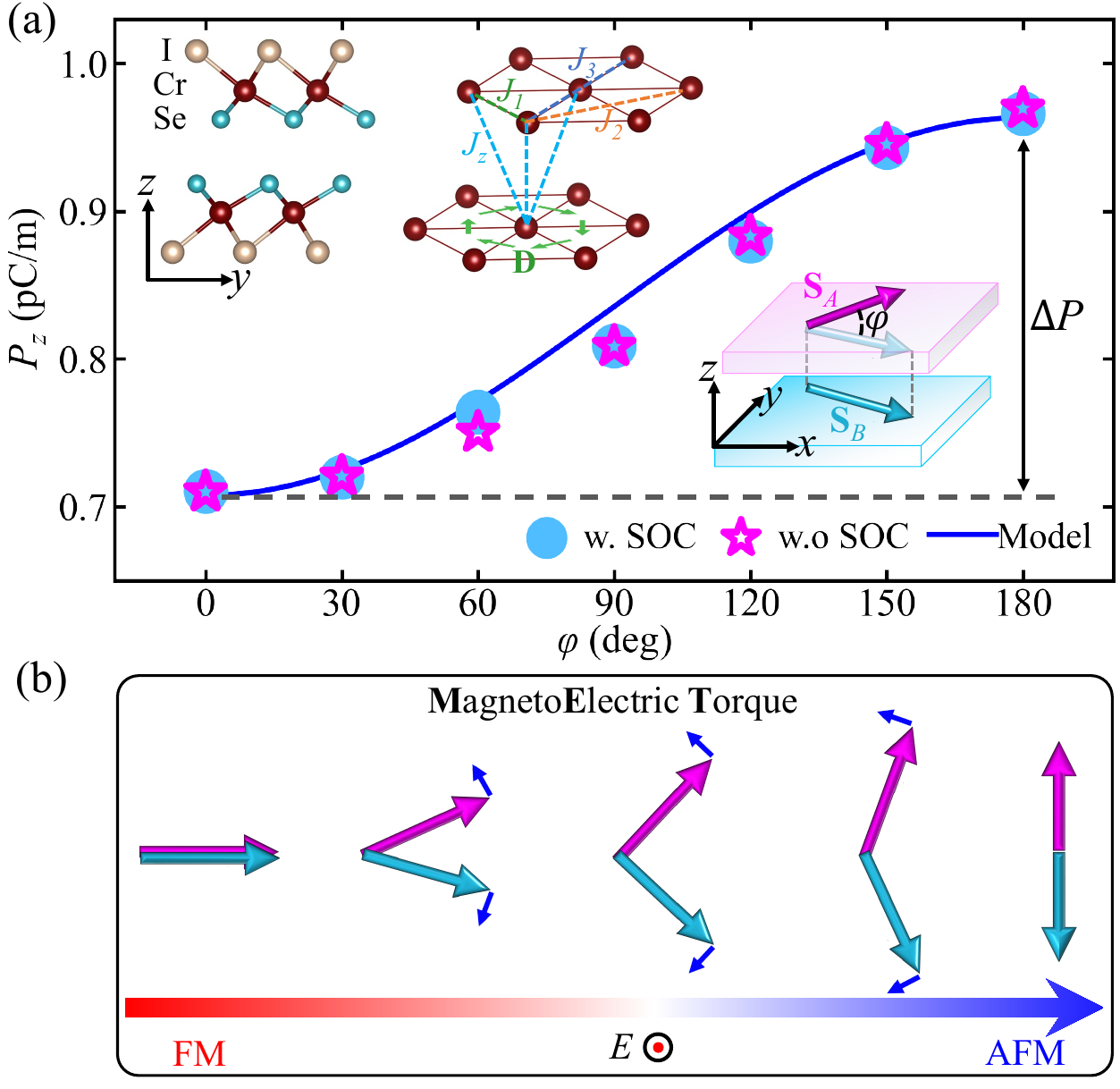}
\caption{(a) The sliding polarization of CrISe bilayer as a function of the angle ($\varphi$) between $\textbf{S}_A$ and $\textbf{S}_B$. Dots (stars): obtained from DFT calculations with (without) SOC, which are almost identical. Curve: analytical fitting using Eq.~\ref{eq1}. Insets: side view of CrISe bilayer, schematics of Heisenberg-type exchanges ($J$'s), Dzyaloshinskii-Moriya interaction (DMI) vectors ($\textbf{D}$'s), and spin angle ($\varphi$) between layers. (b) Schematic diagram of the MET (small blue arrows) on spins (large arrows) in the $xy$-plane, generated by $E$ along the $z$ axis.}
\label{fig1}
\end{figure}

First, the structural and magnetic properties of CrISe bilayer are investigated. For each CrISe monolayer, the magnetic Cr cations form a hexagonal lattice, which is sandwiched by two sheets of nonmagnetic anions (I and Se). Then, the energy-favored stacking sequence of CrISe bilayer is found to be I-Se--Se-I with sliding [space group $P3m1$, as shown in the inset of Fig.~\ref{fig1}(a)] and its dynamic stability is confirmed [see Figs.~S1-S2 in SM~\cite{SMp3} for more details]. Unless otherwise specified, all the following discussions are based on this structure. For CrISe bilayer, an indirect band gap $\sim0.8$ eV is obtained with the Heyd-Scuseria-Ernzerhof (HSE06) hybrid functional~\cite{Heyd2003TJoCP,Heyd2006TJoCP} and the insulation is maintained under electric field $E=\pm0.2$ V/\AA{} [the positive (negative) sign means that $E$ is applied in the direction of $+z$ ($-z$)]. The magnetic ground state is verified to be FM with strong intralayer and moderate interlayer magnetic couplings. Considering the uniform orientation of spins in each layer due to the strong intralayer ferromagnetism, in the following discussions, we use $\textbf{S}_A$ and $\textbf{S}_B$ to represent the unified vectors of spins in the top and bottom layers. $\textbf{S}_A$ and $\textbf{S}_B$ prefer to stay in the $xy$ plane due to a strong easy-plane anisotropy. More details of the band structures and magnetic properties of CrISe bilayer can be found in Figs.~S3-S7 in SM~\cite{SMp3}.

Second, the polarization ($P$) of CrISe bilayer with FM ground state is calculated via density functional theory (DFT), which is normal to the $xy$ plane with $P_z=0.71$ pC/m while its in-plane component is zero due to the $C_{3v}$ rotational symmetry. Here $P_z$ mainly originates from the distortion of electron clouds instead of ionic perpendicular displacements, as in other sliding ferroelectrics~\cite{Wu2021PNAS} (see Fig.~S8 in SM~\cite{SMp3}). Then, as shown in Fig.~\ref{fig1}(a), the two-dimensional $P_z$ depends on the angle ($\varphi$) between $\textbf{S}_A$ and $\textbf{S}_B$, which can be well fitted as:
\begin{eqnarray}
P_z=P_0 - \lambda(\textbf{S}_A\cdot \textbf{S}_B) = P_0 - \lambda \cos\varphi.
\label{eq1}
\end{eqnarray}
Here, $P_0=0.838$ pC/m and $\lambda=0.128$ pC/m is the magnetoelectric coupling coefficient. This magnetoelectricity is almost SOC-independent ($P_0=0.840$ pC/m and $\lambda=0.130$ pC/m without SOC), which can be attributed to the exchange striction~\cite{Sergienko2006PRL}. Different from the one based on the spin-lattice coupling~\cite{Choi2008PRL}, the exchange striction here is dominated by the distortion of electron clouds [Figs.~S9(a)-(c) in SM~\cite{SMp3}]. Similar cases have also been reported in the type-II multiferroic $o$-HoMnO$_3$ and Hf$_2$VC$_2$F$_2$~\cite{Picozzi2007PRL,Zhang2018JotACS}. It should be noted that the broken reversal symmetry along the $z$ axis is the key ingredient to induce this magnetoelectricity, while the sliding ferroelectricity just plays the role of symmetry breaking. The magnetoelectricity also appears in the polar Se-I--Se-I stacking mode even without sliding [Fig.~S9(e) in SM~\cite{SMp3}].

According to Eq.~\ref{eq1}, there is a remarkable polarization difference $\Delta P_z$ between interlayer FM ($\varphi = 0^\circ$) and antiferromagnetic (AFM) ($\varphi = 180^\circ$) states. Thus, when a perpendicular electric field is applied, an extra energy difference between interlayer FM and  AFM orders emerges as $\Delta \varepsilon_E = E_z\Delta P_z$, which can compensate the interlayer magnetic coupling and induce torques acting on spins, as illustrated in Fig.~\ref{fig1}(b). Thereby, the magnetization can be switched from 1 (FM) to 0 (AFM) by electric field. Such a torque depends on the coupling between the interlayer spin order and the polarization, and thus can be coined as MET, which is a unique feature of the polar vdW magnetic bilayers.

The spin model Hamiltonian of CrISe bilayer can be written as:
\begin{eqnarray}
\nonumber H=&&-\sum_{<i,j>}[J_{ij}\textbf{S}_i\cdot\textbf{S}_j+\textbf{D}_{ij}\cdot(\textbf{S}_i\times\textbf{S}_j)]\\
&&+K\sum_i(S_i^z)^2-E_zP_z.
\label{eq2}
\end{eqnarray}
Here, the first two terms are the interactions between magnetic Cr sites $i$ and $j$. $J$'s denote the Heisenberg-type exchange coefficients between the intralayer first, second, and third neighbors ($J_1$, $J_2$, and $J_3$) and interlayer nearest neighbors ($J_z$). $\textbf{D}$ denotes the bond-dependent DMI vector (only the intralayer 1st nearest neighbor DMI is considered, since it is typically much weaker than $J$'s considering its SOC origin). More information of $J$'s and $\textbf{D}$ can be found in the inset of Fig.~\ref{fig1}(a). The third term is the single-site magnetocrystalline anisotropy, with a negative (positive) $K$ for easy axis (easy plane). The last term describes the electrostatic energy and the value of $P_z$ can be estimated using Eq.~\ref{eq1}.

At zero temperature with uniform spins in each layer, the interlayer coupling energy (per spin pair) can be analytically written as a function of $\varphi$: 
\begin{equation}
\varepsilon=-J_z\textbf{S}_A\cdot\textbf{S}_B+\frac{1}{3}E \lambda \textbf{S}_A\cdot \textbf{S}_B=(-J_z+\frac{1}{3}E\lambda)\cos\varphi,
\label{eq3}
\end{equation}
where the coefficient $1/3$ appears due to the three nearest interlayer neighbors. Thus, the energy difference ($\Delta \varepsilon$) between the interlayer FM and AFM orders is $2(-J_z+\frac{1}{3}E\lambda)$. If $E\lambda>3J_z$, the interlayer spin order can be switched from the FM to AFM states by electric field. The critical electric field $E_c$ is obtained as $3J_z/\lambda=3.6$ V/\AA{} for the pristine bilayer. The value of $E_c$ can be finely tuned to be smaller: e.g., $0.014$ V/\AA{}, by reducing $J_z$ via a biaxial tensile strain of $3.75\%$. Such an electric-field-controlled FM-AFM switching is confirmed in DFT calculations [Fig.~S10(d) in SM~\cite{SMp3}].

For better performance, all following model simulations are based on the strain condition of $3.75\%$ with all model parameters listed in Table~\ref{tab1}, extracted from DFT calculations (see SM3 in SM for more details~\cite{SMp3}). With these parameters, the spin model can give the FM ground state with all spins mostly lying in the $xy$ plane, despite the nonzero $\textbf{D}$.

\begin{table}
\centering
\caption{Magnetic and magnetoelectric coefficients extracted from DFT calculations under a strain of $3.75\%$. $P_0$ and $\lambda$ are in units of pC/m, and others are in units of meV. $D$ is the amplitude of $\textbf{D}$.}
\setlength{\tabcolsep}{1.38mm}
\begin{tabular*}{0.48\textwidth}{lcccccccc}
\hline\hline
Layer & $J_1$ & $J_2$ & $J_3$ & $D$ & $K$ & $J_z$ & $P_0$ & $\lambda$\\
\hline
Top & 28.04 & 0.38 & -0.74 & -2.28 & 1.96 & \multirow{2}{*}{0.005} & \multirow{2}{*}{0.56} & \multirow{2}{*}{0.18} \\
Bottom & 28.94 & 0.59 & -0.53 & 2.48 & 2.04 &  &  &  \\
\hline\hline
\end{tabular*}
\label{tab1}
\end{table}

Then, the atomistic Landau-Lifshitz-Gilbert (LLG) equation is employed to simulate the spin dynamics of magnetization switching \cite{Evans2014JoPCM,Lifshitz1935PZS,Gilbert2004IToM}:
\begin{eqnarray}
	\frac{\partial \textbf{S}_i}{\partial t} = -\frac{\gamma}{(1+\alpha^2)\mu_{s}}[\textbf{S}_i\times \textbf{f} + \alpha \textbf{S}_i\times(\textbf{S}_i\times \textbf{f})],
	\label{eq4}
\end{eqnarray}
where $\gamma$ is the gyromagnetic ratio, $\mu_{s} = 3\mu_{B}$ is the magnetic moment of Cr$^{3+}$ ion with $\mu_{B}$ being the Bohr magneton, $\alpha$ is the Gilbert damping coefficient, and $\textbf{f} = -\partial H/\partial \textbf{S}_i$ is the effective field.

\begin{figure}
\centering
\includegraphics[width=0.48\textwidth]{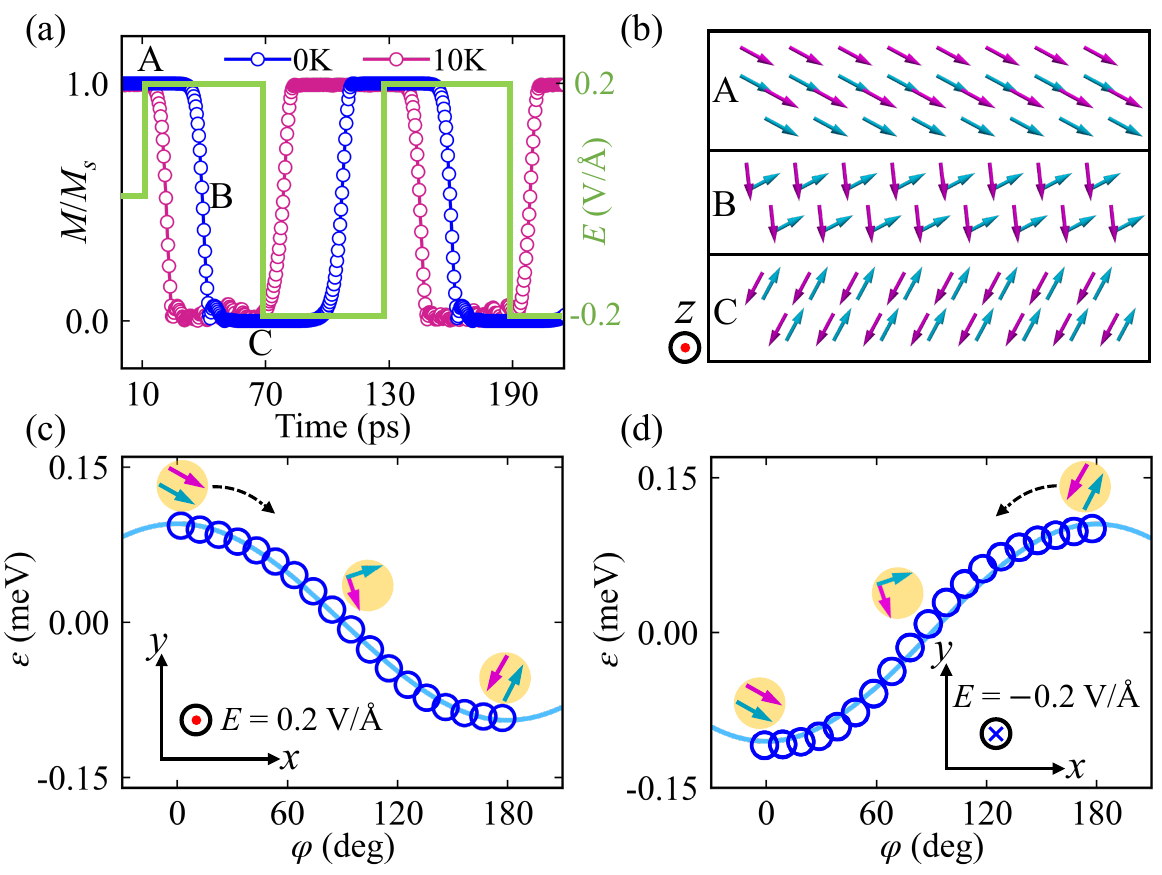}
\caption{The atomistic simulations and an analytic description of the magnetization ($M$) switching. (a) Evolutions of the normalized magnetization over time in the atomistic simulations at $0$ K and $10$ K. $M_s$: the saturation magnetization, i.e., $3$ $\mu_{\rm B}$/Cr. Green curve: $E$ along the $z$ axis. (b) Corresponding spin textures at points A, B, C in (a). (c-d) The analytic (curves) and simulated (circles) energy ($\varepsilon$) as a function of $\varphi$ with (c) $E = 0.2$ V/\AA{} and (d) $E = -0.2$ V/\AA{}. In (b-d), spins in different layers are distinguished by colors.}
\label{fig2}
\end{figure}

Starting from the FM ground state, the zero-temperature simulation result is shown in Fig.~\ref{fig2}(a). At $10$ ps, $E = 0.2$ V/\AA{} is applied. After a retardation of $\sim25$ ps, the magnetization fastly switches to $0$ within $\sim8.0$ ps. When the direction of $E$ is reversed ($E=-0.2$ V/\AA), the magnetization is switched back to $1$ after a similar retardation. Corresponding spin textures during this process [points A, B, C  in Fig.~\ref{fig2}(a)] are shown in Fig.~\ref{fig2}(b). Noting that the electric field here is comparable to those applied on BiFeO$_3$ thin films (experiments)~\cite{Ji2019N,Wang2018NCb} and $M_3X_8$ monolayers~\cite{Xie2025PRL}, and is about one order of magnitude smaller than that in NiI$_2$ bilayer~\cite{Bennett2024PRL}. The finite-temperature simulation at $10$ K is also done, as compared in Fig.~\ref{fig2}(a). The thermal fluctuations can speed up the starting retardation, without obvious negative effects to the switching. In fact, this switching can persist up to $\sim170$ K, noting that its FM Curie temperature is $\sim230$ K (Figs.~S11 and S12 in SM~\cite{SMp3}). The underlying mechanism is shown in Figs.~\ref{fig2}(c) and \ref{fig2}(d). The electric field tunes the energy balance between the FM and AFM states, as confirmed in the atomistic simulations as well as analytical Eq.~\ref{eq3}. According to the energy profiles, $\varepsilon$ changes slowly around $\varphi = 0^\circ$ and $180^\circ$ as the origin of retardation before switching, but much faster near $\varphi = 90^\circ$.

Here the electric-field-driven magnetization switching is in the scale of $10$ picosecond, $\sim10^2$ times faster than the approach based on ion displacements in multiferroics like Co(MoTe$_2$)$_2$~\cite{Yu2024PRL} and CuCrP$_2$Se$_6$~\cite{Yu2023PRL}. This is mainly attributed to the much faster dynamics of electrons than ions, a natural superiority. Besides, the theoretical value of energy consumption for a single switching is estimated to be $\sim2\times10^{-20}$ J/($100$ nm$^2$), $\sim10^5-10^6$ times lower than the STT and SOT technologies~\cite{Dieny2020}, and even $\sim10^2$ times lower than the conceived magnetoelectric spin-orbit (MESO) device~\cite{Manipatruni2018N}.

\begin{figure}
	\centering
	\includegraphics[width=0.48\textwidth]{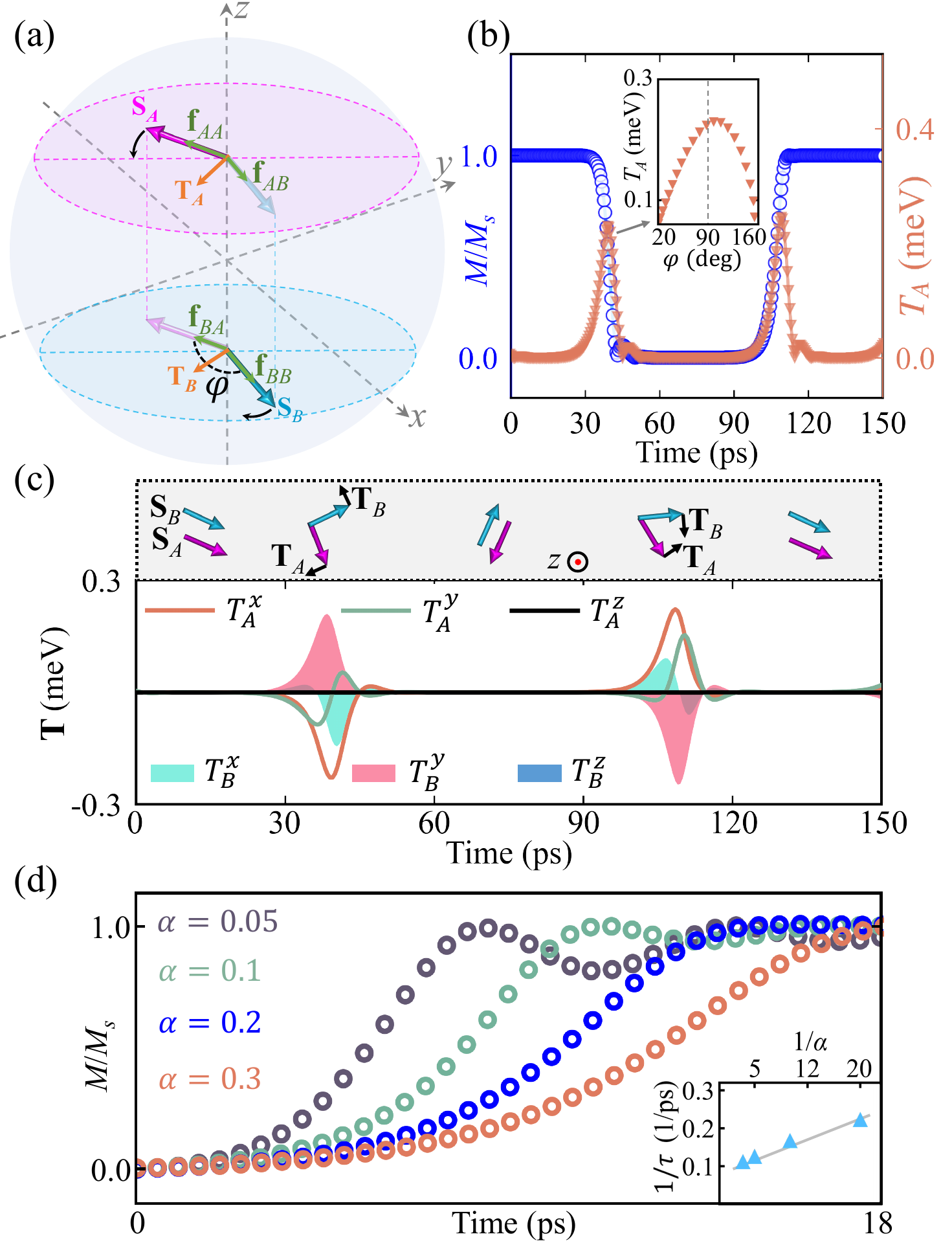}
	\caption{Role of MET in the magnetization switching. (a) $\textbf{S}_A$ ($\textbf{S}_B$) is confined in the easy plane. $\textbf{f}_{AA}$ and $\textbf{f}_{AB}$ ($\textbf{f}_{BB}$ and $\textbf{f}_{BA}$) are the effective fields acting on $\textbf{S}_A$ ($\textbf{S}_B$) from intra- and interlayer interactions. $\textbf{T}_A$ ($\textbf{T}_B$) is the MET acting on $\textbf{S}_A$ ($\textbf{S}_B$). (b) Evolution of magnetization and $T_A$ in the atomistic simulation. Inset: $T_A$ as a funtion of $\varphi$. (c) Evolution of three components of $\textbf{T}_A$ and $\textbf{T}_B$. The corresponding orientations of spins and torques are also illustrated. The electric field in (b-c) is the same as that in Fig.~\ref{fig2}(a). (d) Magnetization switching simulated with different $\alpha$'s under $E = -0.2$ V/\AA{}. Inset: the $M$-switching frequency ($1/\tau$) as a function of $1/\alpha$.}
	\label{fig3}
\end{figure}

To further understand this dynamics, the driving force, i.e., MET, will be illustrated as follows. As shown in Fig.~\ref{fig3}(a), $\textbf{S}_A$ ($\textbf{S}_B$) is mostly lying in the $xy$ plane due to the magnetocrystalline anisotropy. The effective field on $\textbf{S}_A$ contains two parts: $\textbf{f}_{AA}$ from intralayer interactions and $\textbf{f}_{AB}$ from interlayer interactions.
On one hand, due to the in-plane FM order in each monolayer, both the intralayer $\textbf{S}_i\times\textbf{S}_j$ and $S^{z}$ in Eq.~\ref{eq2} are zero in the ideal limit, and thus $\textbf{f}_{AA}=\partial \sum_{j} J_{ij}(\textbf{S}_{A,i} \cdot \textbf{S}_{A,j})/\partial \textbf{S}_{A,i}$. Thus, $\textbf{f}_{AA}$ is in the same direction as $\textbf{S}_A$. Thereby, the torque from intralayer spins $\textbf{T}_{AA} = -\textbf{S}_A \times \textbf{f}_{AA} - \alpha\textbf{S}_A \times (\textbf{S}_A \times \textbf{f}_{AA})= 0$. On the other hand, $\textbf{f}_{AB} = \partial \sum_j [J_z\textbf{S}_{A,i} \cdot \textbf{S}_{B,j} + E (P_0 - \lambda \textbf{S}_{A,i} \cdot \textbf{S}_B)/3]/\partial \textbf{S}_{A,i}= (3J_z - E \lambda) \textbf{S}_B$, in the same direction as $\textbf{S}_B$. Then the MET from interlayer interaction $\textbf{T}_{AB}=-\textbf{S}_A \times \textbf{f}_{AB} - \alpha\textbf{S}_A \times (\textbf{S}_A \times \textbf{f}_{AB})$. Since the direction of $\textbf{S}_A \times \textbf{f}_{AB}$ is perpendicular to the $xy$ plane but $\textbf{S}_A$ is confined within this plane, the first term doesn't play a dominant role (see SM4 and Fig.~S13 in SM~\cite{SMp3} for more details). Thereby, the MET of $\textbf{S}_{A}$ can be simplified as:
\begin{eqnarray}
\textbf{T}_{A} =  -\alpha\textbf{S}_A \times (\textbf{S}_A \times \textbf{f}_{AB}) = \alpha\hat{\textbf{e}}(3J_z - E\lambda)\mathrm{sin}\varphi,
\label{eq5}
\end{eqnarray}
where $\hat{\textbf{e}}$ is a unit vector in the $xy$ plane and perpendicular to $\textbf{S}_A$. Noting that $\textbf{T}_B$ can be obtained with the same method. According to Eq.~\ref{eq5}, the analytic maximums of $T_A$ (i.e., $|\textbf{T}_A|$) should locate at $\varphi = 90^\circ$. 
To verify this point, the numerical $T_A$ from the atomistic simulation is shown in Fig.~\ref{fig3}(b) as a function of time or $\varphi$. There are indeed sharp peaks around $\varphi \sim 90^\circ$. Furthermore, three components of $\textbf{T}_B$ (and $\textbf{T}_A$) are shown in Fig.~\ref{fig3}(c). The simulated MET-driven magnetization switching confirms the scenario of Fig.~\ref{fig1}(b).

Although here the CrISe bilayer is not a typical AFM system, it still exhibits ultrafast dynamics with THz frequency. To clarify this point, we assume the easy-plane anisotropy is strong enough to suppress the out-of-plane vibration (also confirmed in Fig. S14), thus the analytic solution for magnetization switching can be derived with a characteristic time (see SM4 in SM~\cite{SMp3} for more details): $\tau =\pi\alpha\mu_s/[\gamma(E\lambda - 3J_z)]$. The emergent relationship $\tau\varpropto\alpha$ indicates a less-dissipative behavior, which is similar with the spin dynamics in pure AFM systems~\cite{Gomonay2016PRL}. Interestingly, the acceleration of spin dynamics originates from a strong easy-plane anisotropy, which differs from the typical AFM exchange enhancement mechanism, rendering for a new route to realize ultrafast dynamics. With $\alpha = 0.2$, the analytic characteristic time yields an approximate value of $\tau\sim7.2$ ps, which is on the same order of magnitude as the simulation data ($\sim8.0$ ps). Given that $\tau$ is an approximate value under the condition $K\to +\infty$, it's reasonable for $\tau$ to be smaller than the simulated one. Additionally, simulations conducted with different $\alpha$'s are presented in Fig.~\ref{fig3}(d). These results indicate that the switching frequency is nearly proportional to $1/\alpha$ [as shown by the inset of Fig.~\ref{fig3}(d)], which is consistent with our analytic solution. Typically, the Gilbert damping coefficient should be much smaller than $0.1$, thus the ultrafast dynamics should be robust in practice. 

\begin{figure}
\centering 
\includegraphics[width=0.48\textwidth]{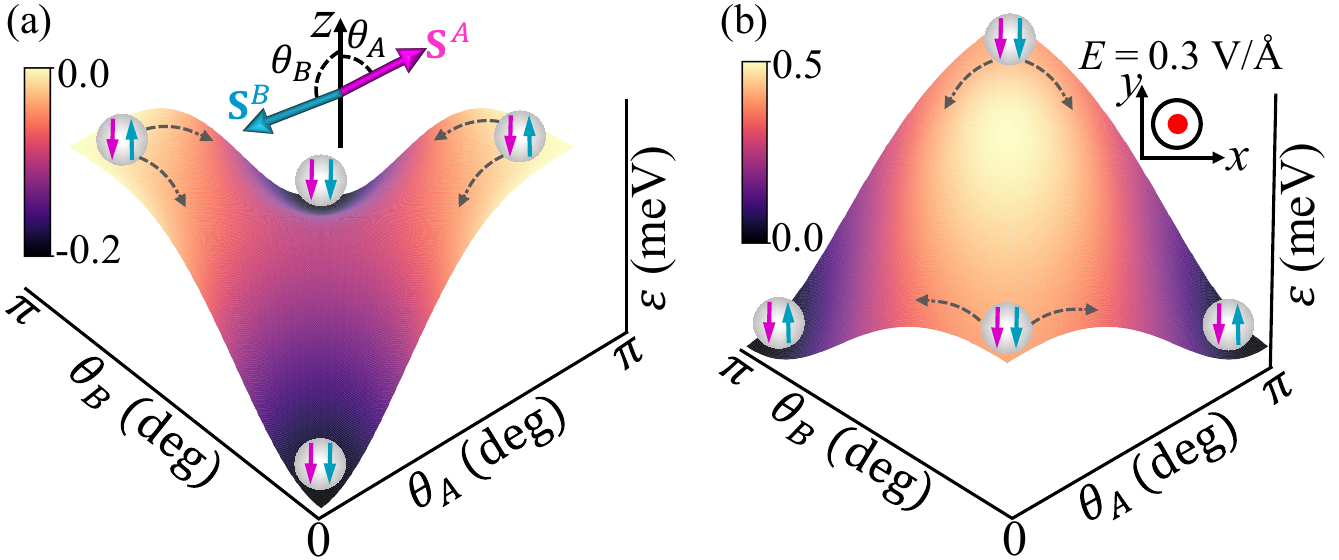}
\caption{The analytic energy ($\varepsilon$) surface of an easy-axis magnet (a) without and (b) with an electric field of 0.3 V/\AA{}. $\theta_A$ ($\theta_B$) is the polar angle of $\textbf{S}_A$ ($\textbf{S}_B$). The value of $\varepsilon$ is represented by the color map. The magnetization switching paths are shown by the dashed grey arrows.}
\label{fig4}
\end{figure}

All above discussions regarding the MET-driven magnetization switching were based on the easy-plane magnet. However, this MET mechanism also works for those magnets with easy-axis magnetization. To avoid verbosity while retaining generality, we give a lucid analytic demonstration of the MET-driven magnetization switching in easy-axis magnets. Similar to CrISe bilayer, we assume the easy-axis magnet with a strong intralayer and a moderate interlayer ($J_1 = 10$ meV, $J_2 = J_3 = 0$ meV and $J_z = 0.1$ meV) FM order. The magnetocrystalline anisotropy coefficient is $-$0.05 meV for both top ($K_A$) and bottom ($K_B$) layers. $P_0 = 1.0$ pC/m and $\lambda = 0.35$ pC/m. The polar angle of spin in the top (bottom) layer is $\theta_A$ ($\theta_B$) as shown in Fig.~\ref{fig4}(a).

Once $\theta_A$ and $\theta_B$ are given, $\textbf{S}_A$, $\textbf{S}_B$, and $z$ axis always tend to be coplanar to maximize $\textbf{S}_A\cdot\textbf{S}_B$ (i.e., to minimize the interlayer exchange energy). Then, the energy related to spin dynamics ($\varepsilon$, per $\textbf{S}_A$-$\textbf{S}_B$ pair) can be written as:
\begin{eqnarray}
	\varepsilon =&& -J_z(\textbf{S}_A\cdot \textbf{S}_B) + E\lambda(\textbf{S}_A\cdot \textbf{S}_B)/3 \nonumber  \\
	&& + K_A\cdot (\textbf{S}_A^z)^2 + K_B\cdot (\textbf{S}_B^z)^2 \nonumber  \\
	= &&( -J_z + E\lambda/3)(\sin{\theta_A}\sin{\theta_B} + \cos{\theta_A}\cos{\theta_B})\nonumber  \\
	&& + K_A\cos^2{\theta_A} + K_B\cos^2{\theta_B}.
	\label{eq6}
\end{eqnarray}

According to Eq.~\ref{eq6}, the analytic energy surface of $\varepsilon$ is plotted in Fig.~\ref{fig4} with the energy of interlayer AFM order as the reference value. As shown in Fig.~\ref{fig4}(a), without electric field, the saddle-like energy surface indicates the FM ground state, while by applying an electric field of $0.3$ V/\AA{}, the energy surface becomes a mountain-like one with the interlayer AFM ground state. Thus, the magnetization of easy-axis polar bilayer magnets can also be switched by electric field via the MET. The underlying mechanism is similar to that in CrISe bilayer [Fig.~\ref{fig2}(c-d)], i.e., the energy balance between FM and AFM states can be finely tuned by the electric field, and thus a MET will drive the magnetization switching. Of course, a relative larger $E$ is needed to overcome the energy barrier from $K$ during the spin dynamics.

In summary, an alternative mechanism named MET is proposed to switch magnetization in polar vdW magnetic bilayer by electric field. This MET is based on the converse magnetoelectricity, which is independent of the spin-orbit coupling. Benefitting from the ultrafast dynamics of electron cloud, the magnetization switching can be realized within $10$ picoseconds. In principle, this kind of magnetoelectricity generally exists in polar magnetic vdW homo- and heterostructures (Table S1 in SM~\cite{SMp3}). Our work not only introduces theoretical insights into the electric-field control of magnetism but also provides guidance for the design of energy-efficient spintronic devices.

\begin{acknowledgments}
This work is supported by National Natural Science Foundation of China (Grant No. 12325401, No. 12274069, No. 12574090, No. 124B2064, and No. 12504123), Jiangsu Funding Program for Excellent Postdoctoral Talent and the
Postdoctoral Fellowship Program China of CPSF (Grant
No. GZB20250766), the Postgraduate Research \& Practice Innovation Program of Jiangsu Province (Grant No. KYCX24\_0362), and the SEU Innovation Capability Enhancement Plan for Doctoral Students (Grant No. CXJH\_SEU 25002), the Big Data Computing Center of Southeast University, and the Center for Fundamental and
Interdisciplinary Sciences of Southeast University.
\end{acknowledgments}

%\bibliography{F:/job/JabRef/physics}
\bibliography{physics}

\onecolumngrid
\vspace*{0.5cm}
\begin{center}
	{\large \bfseries End Matter}
\end{center}
\vspace*{0.5cm}

\twocolumngrid
{\it EM1: The key distinctions between our MET and those of previous ones.}

\begin{figure}[htbp]
\centering
\includegraphics[width=0.4\textwidth]{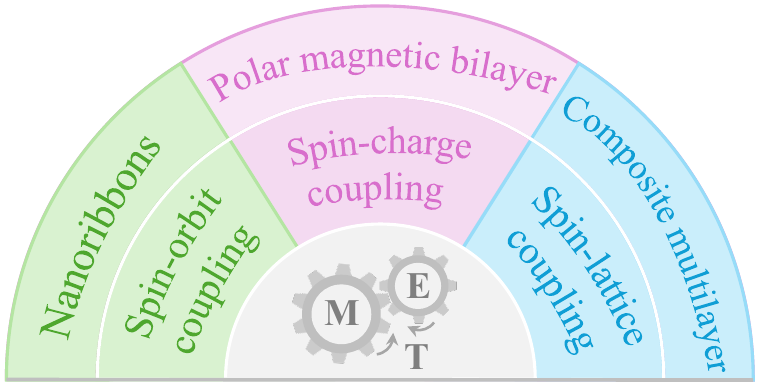}
\caption{A schematic illustration of the key distinctions between our MET (middle) and those of previous ones \cite{Sousa2021PRR,Zheng2017CPB,Xing2013SaAAP,Xing2011JoAP}.}
\label{fig5}
\end{figure}

As shown in Fig.~\ref{fig5}, different from the METs proposed in previous works, which are mainly driven via the spin-orbit coupling~\cite{Sousa2021PRR}, or the spin-lattice coupling~\cite{Zheng2017CPB,Xing2013SaAAP,Xing2011JoAP}, or a combination of both~\cite{Nan2018AFM}, our MET is dominated by the spin-charge coupling (i.e., the polarization difference between interlayer AFM and FM orders), which remains robust even without SOC or lattice distortion. Considering that the SOC is rather weak in most materials and the atomic motions are much slower than that of electrons, our MET can be more advantageous regarding its stronger magnetoelectric effect and faster spin dynamics with THz frequency.

Moreover, the material systems are different. In prior works, the target material systems are nanoribbons with geometric confinement~\cite{Sousa2021PRR}, or composite multilayers composed of piezoelectric layers and magnets~\cite{Zheng2017CPB,Xing2013SaAAP,Xing2011JoAP}. While our MET generally exists in intrinsic polar magnetic bilayers (homo- and heterostructures).

In addition, the MET in Ref~\cite{Sousa2021PRR} tends to create noncollinear spin polarization, and those of Refs~\cite{Zheng2017CPB,Xing2013SaAAP,Xing2011JoAP} will induce a voltage signal for magnetic field detection. While our MET can lead to a fast switching of the whole magnetization between 0 and 1. Thus the resulting effects are also different.

{\it EM2: The differences between our work and Ref.~\cite{Bennett2024PRL}.}

Although the magnetization switching in Ref.~\cite{Bennett2024PRL} is also based on the polarization difference between different interlayer magnetic orders, we extend the practical range of this mechanism to cover both the easy-axis and easy-plane magnets in various polar magnetic bilayers (Table S1).

Different from the easy-axis anisotropy that needs to be overcome during magnetization switching~\cite{Bennett2024PRL}, the easy-plane anisotropy in our work results in nearly zero energy barrier of spin rotation within the $xy$ plane and thus enable a more accessible magnetization switching.

Furthermore, the focuses of these two works are very different. Ref.~\cite{Bennett2024PRL} focuses on the stacking-engineered multiferroic order and magnetoelectric coupling. While our work focuses on the energy-efficient magnetization switching with THz frequency, thus rendering a new route to realize ultrafast spin dynamics.

\end{document}